\documentclass[prb,superscriptaddress,twocolumn]{revtex4-1}
\usepackage[pdftex]{graphicx}
\usepackage{amsmath}
\usepackage{chngcntr}

\draft 

\begin{document}

\title{Quantum Circuit Distillation and Compression}

\author{Shunsuke Daimon}
\email[E-mail: ]{daimon.shunsuke@qst.go.jp}

\affiliation{Department of Applied Physics, The University of Tokyo, Tokyo 113-8656, Japan.}
\affiliation{Department of Applied Physics and Physico-Informatics, Keio University, Yokohama 223-8522, Japan.}
\affiliation{Institute for AI and Beyond, The University of Tokyo, Tokyo 113-8656, Japan.}
\affiliation{Quantum Materials and Applications Research Center, National Institutes for Quantum Science and Technology, Tokyo 152-8550, Japan.}

\author{Kakeru Tsunekawa}
\affiliation{Department of Applied Physics, The University of Tokyo, Tokyo 113-8656, Japan.}

\author{Ryoto Takeuchi}
\affiliation{Department of Applied Physics, The University of Tokyo, Tokyo 113-8656, Japan.}

\author{Takahiro Sagawa}
\affiliation{Department of Applied Physics, The University of Tokyo, Tokyo 113-8656, Japan.}

\author{Naoki Yamamoto}
\affiliation{Department of Applied Physics and Physico-Informatics, Keio University, Yokohama 223-8522, Japan.}

\author{Eiji Saitoh}
\affiliation{Department of Applied Physics, The University of Tokyo, Tokyo 113-8656, Japan.}
\affiliation{Department of Applied Physics and Physico-Informatics, Keio University, Yokohama 223-8522, Japan.}
\affiliation{Institute for AI and Beyond, The University of Tokyo, Tokyo 113-8656, Japan.}
\affiliation{Institute for Materials Research, Tohoku University, Sendai 980-8577, Japan.}
\affiliation{WPI Advanced Institute for Materials Research, Tohoku University, Sendai 980-8577, Japan.}


\begin{abstract}
Quantum coherence in a qubit is vulnerable to environmental noise. When long quantum calculation is run on a quantum processor without error correction, the noise often causes fatal errors and messes up the calculation. Here, we propose quantum-circuit distillation to generate quantum circuits that are short but have enough functions to produce an output almost identical to that of the original circuits. The distilled circuits are less sensitive to the noise and can complete calculation before the quantum coherence is broken in the qubits. We created a quantum-circuit distillator by building a reinforcement learning model, and applied it to the inverse quantum Fourier transform (IQFT) and Shor’s quantum prime factorization. The obtained distilled circuit allows correct calculation on IBM-Quantum processors. By working with the quantum-circuit distillator, we also found a general rule to generate quantum circuits approximating the general $n$-qubit IQFTs. The quantum-circuit distillator offers a new approach to improve performance of noisy quantum processors.
\end{abstract}

\maketitle

\par
\section{Introduction}
Quantum calculation has been performed on gate-type quantum processors by designing a sequence of quantum logic gates called a quantum circuit{\cite{ladd2010quantum,alexeev2021quantum}}. Numerous algorithms based on quantum circuits have theoretically been proposed to realize quantum computation{\cite{nielsen2010quantum,montanaro2016quantum}}. However, existing quantum processors are too noisy and not yet ready to perform these quantum algorithms properly{\cite{preskill2018quantum,corcoles2019challenges}}. The noise causes errors in running quantum circuits and prevents the circuits from working properly{\cite{chow2012universal,willsch2017gate}}, and the quantum processors cannot even perform simple quantum algorithms.\par
Here, we propose a quantum-circuit distillator to generate a quantum circuit that is short but has enough functions to yield almost the same output distribution as that of a long original quantum circuit. We show that the quantum-circuit distillator enables existing quantum processors to work properly by reducing errors effectively by demonstrating the distillation of the IQFT and Shor’s integer factorization circuits. By extrapolating the outputs from the distillator, we also found approximated general $n$-qubit IQFTs for any qubit number $n$.\par
\section{Results and Discussion}
\begin{figure*}
\includegraphics{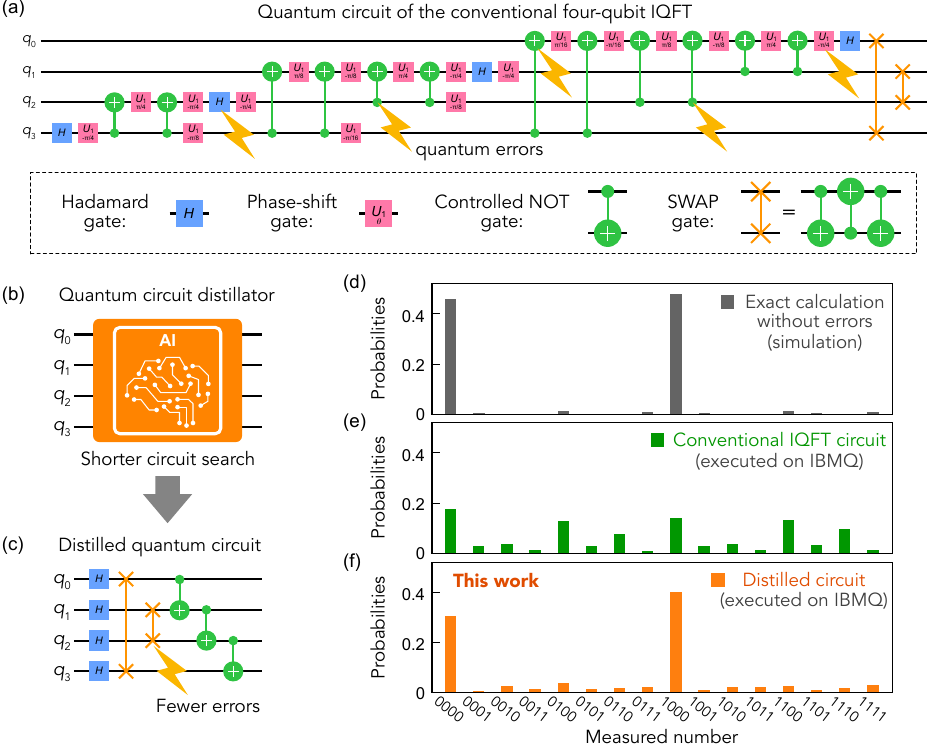}
\caption{\label{fig1}(a) A quantum circuit for the inverse quantum Fourier transform (IQFT) for four qubits $q_0$, $q_1$, $q_2$, and $q_3$. The blue, red, green, and orange components represent the Hadamard, phase-shift, controlled NOT, and SWAP gates, respectively. (b) Schematic of a quantum circuit search. (c) A distilled quantum circuit for the four-qubit IQFT generated from the distillator. (d) A probability distribution obtained from the exact IQFT calculation without errors. (e), (f) Probability distributions obtained from the quantum computer IBMQ for the conventional four-qubit IQFT circuit (e) and for the distilled quantum circuit (f).}
\end{figure*}

First, we show a result of the conventional four-qubit IQFT obtained with an existing superconducting quantum computer IBM Quantum (IBMQ){\cite{gambetta2017building,wendin2017quantum,IBMQuantum}}. Figure {\ref{fig1}}(a) shows the conventional quantum circuit for the four-qubit IQFT{\cite{nielsen2010quantum}}. By applying the circuit to an initial quantum state and measuring the final state, the IBMQ outputs a four-digit bitstring. By repeating the measurement, we obtained probability distribution of the bitstrings. Figure {\ref{fig1}}(e) shows an output distribution profile for the four-qubit IQFT obtained from IBMQ. On the other hand, Fig. {\ref{fig1}}(d) shows the exact analytical result of the four-qubit IQFT calculation for the same input. The distribution observed experimentally from IBMQ [Fig. {\ref{fig1}}(e)] is far different from the expected distribution [Fig. {\ref{fig1}}(d)], showing that the conventional quantum circuit is too long to properly work on IBMQ due to the noise.\par
Before showing the details of the distillator developed in the present study, we demonstrate how it works. When our distillator is applied to the four-qubit IQFT [Fig. {\ref{fig1}}(b)], it compresses the conventional IQFT circuit and outputs a distilled brief quantum circuit as shown in Fig. {\ref{fig1}}(c). We can carry out IQFT calculation using the obtained circuit. Figure {\ref{fig1}}(f) shows the probability distribution of the four-qubit IQFT obtained from the distilled quantum circuit for the same input as those in Fig. {\ref{fig1}}(d) and (e). Significantly, the output profile is almost the same as the exact calculation [Fig. {\ref{fig1}}(d)]; the four-qubit IQFT is successfully completed in spite of the noisy qubit environment in IBMQ owing to the distillation. The distilled quantum circuit is four times shorter than the conventional quantum circuit with 36 quantum gates, and the brevity is responsible for the fewer errors and the successful calculation.\par
\begin{figure*}
\includegraphics{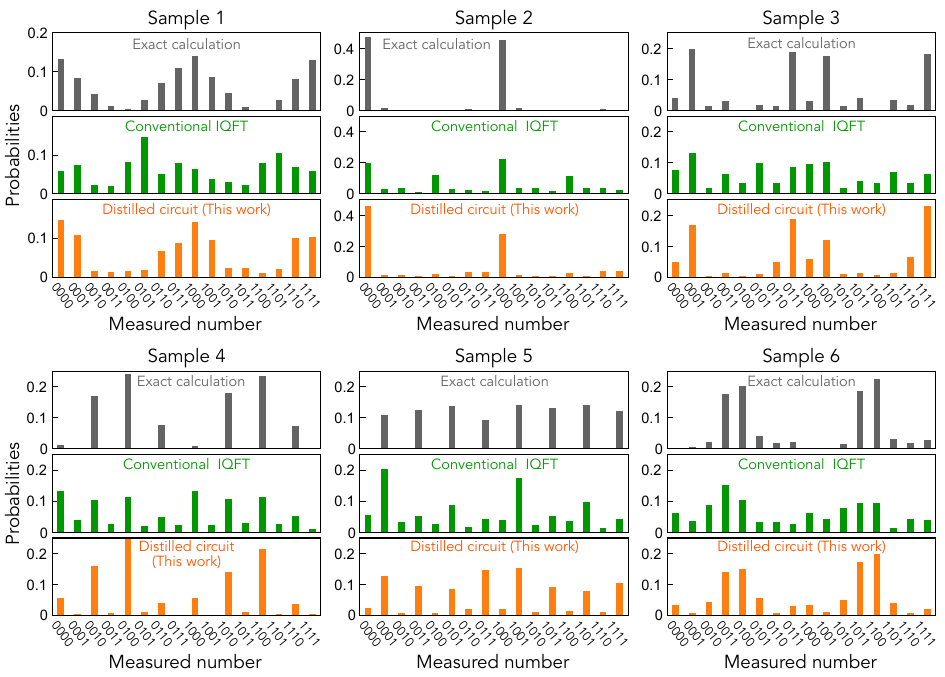}
\caption{\label{fig2} The conventional four-qubit IQFT circuit and the distilled quantum circuit were executed for randomly initialized input states on IBMQ. The probabilities obtained from the exact calculations without errors (the gray bars) and experimental results on IBMQ for the conventional circuit and the distilled quantum circuit (the green and orange bars, respectively) for six different initial states are exemplified.}
\end{figure*}
The main concept of the quantum-circuit distillator is to compress a target quantum circuit into a shorter quantum circuit that well approximates input/output relation of the target quantum circuit. To realize such a distillator, it is necessary to search an ideal quantum circuit from a huge number of combinations of quantum logic gates. We addressed this problem by taking advantage of the remarkable searching abilities of reinforcement learning{\cite{kaelbling1996reinforcement,arulkumaran2017deep}}, known to be suitable for solving optimization problems for finite element combinations, such as quantum gates. To construct a reinforcement learning model for quantum circuit compression, we designed a variable-length quantum circuit search based on the Monte-Carlo tree search in which a deep neural network is used to accelerate the tree search{\cite{silver2017mastering,silver2017mastering2}} (see Appendixes D and E and Figs. {\ref{fig7}} and {\ref{fig8}} for more details). In contrast to the quantum circuit search using fixed-length quantum circuits{\cite{peruzzo2014variational, farhi2014quantum, endo2021hybrid, cerezo2021variational, nakaji2022approximate}}, such as the quantum circuit learning{\cite{mitarai2018quantum}}, our model enables us to search quantum circuits with different lengths and different circuit structures. The deep neural network makes the quantum circuit search efficient by sequentially predicting the most suitable gate to be placed next to the gate array constructed so far in the tree search.\par
\begin{figure*}
\includegraphics{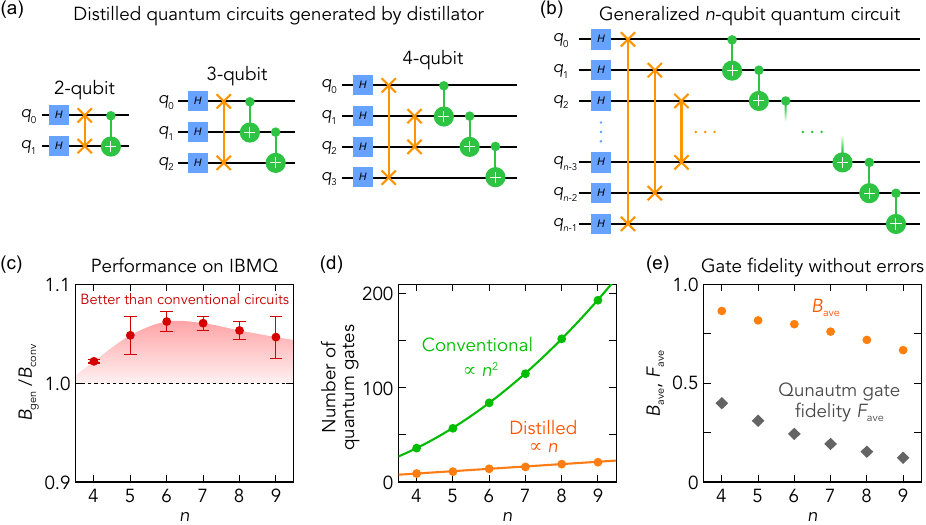}
\caption{\label{fig3}(a) The distilled quantum circuits obtained from the distillator for the two-, three-, and four-qubit IQFTs. (b) The generalized $n$-qubit quantum circuit for the $n$-qubit IQFT. (c) The average performance of the generalized quantum circuits normalized by that of the conventional IQFT circuits, where $B_{\rm{gen}}$ and $B_{\rm{conv}}$ are the averaged Bhattacharyya coefficients of the generalized circuits and the conventional IQFT circuits, respectively. The number of qubits $n$ is changed from 4 to 9. (d) The number of quantum gates used in the generalized circuits and the conventional IQFT circuits. (e) The averaged Bhattacharyya coefficients $B_{\rm{ave}}$ and the quantum gate fidelity $F_{\rm{ave}}$ numerically calculated for the generalized circuits. Here, $B_{\rm{ave}}$ ($F_{\rm{ave}}$) is calculated by comparing the probabilities (unitary matrices) obtained from the exact analytical calculations of the generalized quantum circuits without errors.}
\end{figure*}

We found that the Bhattacharyya coefficient{\cite{fuchs1999cryptographic}} $B$ works well as a reward function in the reinforcement learning used in the quantum-circuit distillation. The Bhattacharyya coefficient $B$ is a classical expression of the quantum fidelity, which gives a measure for the similarity between two quantum circuits based on the probability distributions (see Appendix A for the definition of $B$). $B$ takes a value between 0 and 1, and approaches 1 (0) when the two probability distributions are the same (dissimilar). The quantum gate fidelity{\cite{fuchs1999cryptographic}}, which compares quantum circuits as unitary transformations, has often been used as a reward function in the previous studies on quantum circuit search{\cite{khaneja2005optimal, khatri2019quantum, dalgaard2020global, peng2020simulating, magann2021pulses, fosel2021quantum, moro2021quantum, kimura2022quantum}}. However, we found that it imposes too strong constraints to the search in the present reinforcement learning [see Fig. {\ref{fig3}}(e)]. On the other hand, $B$ only compares probability distributions obtained from quantum circuits and makes it easier to find optimal circuits among a large number of quantum gate combinations.\par
By applying the developed quantum-circuit distillator using the reinforcement learning to the four-qubit IQFT circuit shown in Fig. {\ref{fig1}}(a), we obtained the distilled IQFT circuit shown in Fig. {\ref{fig1}}(c). As we discussed above, the distilled circuit outputs the probability distribution shown in Fig. {\ref{fig1}}(f) for an input, and the distribution is almost the same as the exact calculation shown in Fig. {\ref{fig1}}(d), greatly improved from the original IQFT circuit [Fig. {\ref{fig1}}(e)]. By comparing the probability distributions obtained from the distilled circuit [Fig. {\ref{fig1}}(f)] and the exact calculation [Fig. {\ref{fig1}}(d)], we obtained the $B$ value of 0.91. In contrast, for the probability distribution obtained from the conventional quantum circuit shown in Fig. {\ref{fig1}}(e), $B$ takes a lower value, 0.69, demonstrating the superiority of the distilled circuit.\par
We confirmed the generalization performance of the present reinforcement learning in the circuit distillator; we found that the distilled quantum circuit shown in Fig. {\ref{fig1}}(c) well reproduces the exact calculation of the four-qubit IQFT also for the input quantum states not used for the learning (see Appendix C for the theoretical analysis of the distilled quantum circuit). Figure {\ref{fig2}} shows output probability distributions for several input quantum states prepared by random sequences of quantum gates. All the outputs from the distilled quantum circuit (orange bars in Fig. {\ref{fig2}}) well reproduce the exact calculations (gray bars in Fig. {\ref{fig2}}), while the outputs from the conventional IQFT circuit (green bars in Fig. {\ref{fig2}}) are far from the exact calculations (gray bars in Fig. {\ref{fig2}}).\par
We calculated the average value of the Bhattacharyya coefficient $B_{\rm{ave}}$ for randomly initialized input quantum states; we prepared input states by randomly applying quantum logic gates to the ground state of the quantum bits, and then the distilled quantum circuit is applied to the initial states (see Appendix A and Fig. {\ref{fig5}}). $B_{\rm{ave}}$ for the distilled four-qubit IQFT circuit was estimated to be $\sim$ 0.93, which is greater than that for the conventional circuit 0.90. The result shows again that the distilled quantum circuit with fewer quantum gates generates better outputs than the longer circuit derived from the exact quantum algorithm.\par
By exploiting some outputs from the quantum-circuit distillator, we can infer general quantum circuits which can approximate the $n$-qubit IQFTs for any natural number $n$, as follows. In Fig. {\ref{fig3}}(a), we show distilled circuits obtained from our distillator for the 2-, 3-, and 4-qubit IQFTs. By comparing these distilled quantum circuits shown in Fig. {\ref{fig3}}(a), we found a regularity on the array of the quantum gates: the gates are arranged in the order of Hadamard, SWAP, and CNOT gates in the distilled circuits [compare the distilled 2-, 3- , and 4-qubit IQFTs in Fig. {\ref{fig3}}(a)]. From the observation, we inductively predicted approximate quantum circuits generalized to arbitrary $n$ as shown in Fig. {\ref{fig3}}(b). We found that the generalized quantum circuits actually well approximate the conventional $n$-qubit IQFT even for $n$ greater than four. The averaged Bhattacharyya coefficients for the generalized circuits were found to be greater than that for the conventional IQFT circuits even for $n$ greater than four as shown in Fig. {\ref{fig3}}(c). The result demonstrates that the generalized circuits outperform the conventional IQFT circuits on the existing quantum computer IBMQ. As shown in Fig. {\ref{fig3}}(d), the generalized quantum circuits reduce the number of the gates from $O(n^2)$ to $O(n)$ compared to the conventional IQFTs{\cite{nielsen2010quantum}}. The number of gates is less than that of an approximation circuit $O(n \log n)$ proposed in previous studies{\cite{barenco1996approximate,nam2020approximate}}. The gate number reduction mitigates errors in computation and contributes to the performance of the generalized circuits. Note that the generalized circuits do not approximate the IQFT in terms of unitary transformations while they approximate the output probability distributions. As shown in Fig. {\ref{fig3}}(e), while the output probability distributions show high values of $B_{\rm{ave}}$, the quantum gate fidelity $F_{\rm{ave}}$ takes much lower values for all $n$. If the quantum gate fidelity is used as a reward for the search, as in previous studies{\cite{fosel2021quantum,moro2021quantum}}, the distilled circuit cannot be found. In addition, we have succeeded in analytically justifying the generalized approximate quantum circuits for a special case as described in Appendix C.\par
\begin{figure*}
\includegraphics{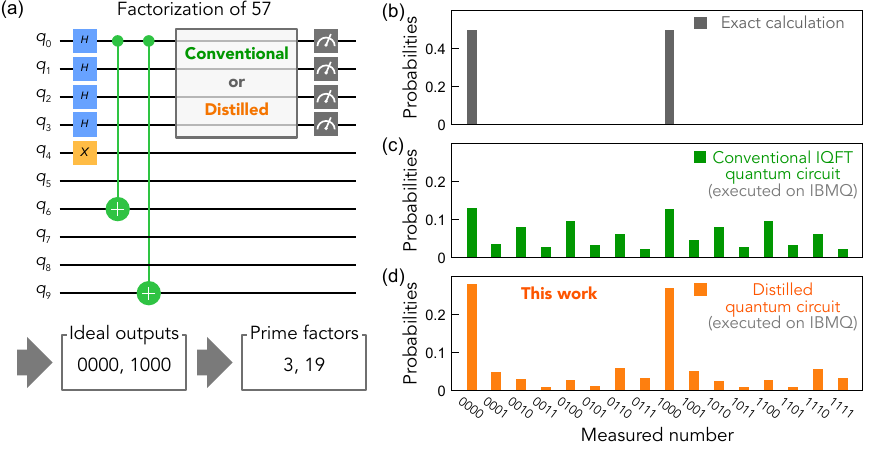}
\caption{\label{fig4}(a) A quantum circuit for the factorization of 57. When the conventional IQFT is chosen in the gray, the quantum circuit solves the order ($r$) finding problem $xr \equiv 1 \pmod {57}$, where we chose $x$ as 37. In this case, the quantum circuit is expected to output “0000” or “1000”, and then the prime factors of 3 and 19 can be derived from the output numbers. The blue, yellow, and green components represent the Hadamard, NOT, and CNOT gates, respectively. (b) Ideal probabilities obtained from the exact analytical calculation without errors. (c), (d) Experimental results for the conventional IQFT circuit (c) and the distilled IQFT circuit (d) obtained from IBMQ.}
\end{figure*}

Finally, we demonstrate the application of the distilled IQFT circuit to executing Shor’s integer factorization{\cite{nielsen2010quantum,ekert1996shor}} of 57. Figure {\ref{fig4}}(a) shows the conventional quantum circuit for the factorization. The quantum circuit ideally outputs 0000 or 1000 [Fig. {\ref{fig4}}(b)] for the input of 57, and the prime factors of 3 and 19 can be derived from the output numbers. When the conventional quantum circuit is run on the quantum computer IBMQ, the output distribution profile [Fig. {\ref{fig4}}(c)] is far different from the exact calculation [Fig. {\ref{fig4}}(b)], and the desired numbers are no longer outputted as shown in Fig. {\ref{fig4}}(c). On the other hand, by replacing the IQFT with the distilled quantum circuit, we obtained almost the same distribution profile as the exact calculation [compare Fig. {\ref{fig4}}(b) and (d)]. The result shows that the distilled quantum circuits enable Shor’s integer factorization on the system.\par
\section{Conclusion}
In summary, we proposed and demonstrated a quantum-circuit distillator to generate quantum circuits less sensitive to noise in existing quantum processors. We applied the quantum-circuit distillator to the four-qubit IQFT and demonstrated that the generated distilled quantum circuit outperforms the conventional IQFT circuit on an existing quantum computer IBMQ. The distilled quantum circuits also let us discover the generalized quantum circuits approximating the general $n$-qubit IQFTs. The result shows a new direction of scientific research, in which humans and machine learning work together to create knowledge. Our method has a wide range of applications, including not only distillation of quantum algorithms such as IQFT, but also data encoding for quantum machine learning. This work will accelerate the realization of valuable quantum computations for practical tasks.
\begin{acknowledgments}
The authors thank N. Yokoi for valuable discussions. This work was supported by ERATO (No. JPMJER1402), CREST (Nos. JPMJCR20C1, JPMJCR20T2) from JST, Japan; Grant-in-Aid for Scientific Research (S) (No. JP19H05600), Grant-in-Aid for Research Activity Start-up (No. JP19K21035), Grant-in-Aid for Transformative Research Areas (No. JP22H05114) from JSPS KAKENHI, Japan. T.S. is supported by JST ERATO-FS (No. JPMJER2204). N.Y. is supported by MEXT Quantum Leap Flagship Program (Nos. JPMXS0118067285, JPMXS0120319794). We acknowledge the use of IBM Quantum services for this work as part of the IBM Quantum Hub at The University of Tokyo and Keio University. The views expressed are those of the authors and do not reflect the official policy or position of IBM or the IBM Quantum team.\par
\end{acknowledgments}
\appendix
\section{Performance estimation of quantum circuits}
Performance of a quantum circuit on a quantum processor is estimated by calculating $B$ between the output probabilities from the quantum circuit $p_{\rm{qc}}$ and the correct answer $p_{\rm{ans}}$ calculated from the ideal IQFT for an initial quantum state $\left|\Psi\right>$. The initial quantum state is prepared by applying a random sequence of quantum gates $g_1,g_2, \cdots,g_m$ to the zero state $\left|0\right>: \left|\Psi\right>=g_m\cdots g_2 g_1 \left|0\right>$, where $g_1,g_2,\cdots,g_m$ are selected from the Hadamard, NOT, Phase-shift, and CNOT gates. Figure {\ref{fig7}}(b) shows an example of quantum gates used for the initialization of 3-qubit quantum states. The number of the quantum gates for the initialization $m$ is set to $4n$ for $n$-qubit quantum circuits in our calculation. Next, the $\left|\Psi\right>$ is transformed by the quantum circuit and measured [Fig. {\ref{fig5}}(b)]. By repeating the measurement, we obtain the probability distribution $p_{\rm{qc}}$ for the input quantum state $\left|\Psi\right>$. At the same time, we numerically calculate the IQFT for the input quantum state $\left|\Psi\right>$ and obtained the probability distribution $p_{\rm{ans}}$ [Fig. {\ref{fig5}}(a)]. Then, $B(p_{\rm{qc}} ,p_{\rm{ans}} )=\sum_{i\in\Omega} \sqrt{p_{\rm{qc}}(i) p_{\rm{ans}}(i)}$ is calculated, where $\Omega$ is the set of the observed binary numbers for the $n$-qubit quantum circuits. By changing the initialization gates and repeating the $B$ calculation for many input quantum states, $B_{\rm{ave}}$ is obtained as the average value of $B$ for the input quantum states.
\renewcommand{\thefigure}{A·\arabic{figure}}
\setcounter{figure}{0}
\begin{figure}
\includegraphics{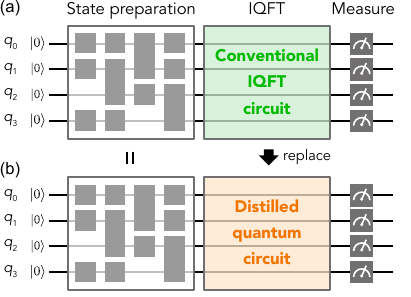}
\caption{\label{fig5}(a), (b) Procedure of performance estimation on a quantum processor for the conventional IQFT circuit (a) and the distilled quantum circuit (b). Firstly, a quantum state is prepared by randomly applying quantum gates to the ground state. Secondly, the prepared quantum state is transformed by the conventional IQFT or distilled quantum circuits. Thirdly, the transformed state is measured. The average performance $B_{\rm{ave}}$ is calculated by averaging the Bhattacharyya coefficient $B$ for many initial quantum states with different initialization quantum gates.}
\end{figure}
\section{Devices and its noise parameters used in the performance estimation}
The calculation results shown in Figs. {\ref{fig1}} and {\ref{fig2}} were obtained from the quantum computer IBMQ Vigo, where the average relaxation times $T_1$ and $T_2$, and readout error rate are about $8 \times 10^1$ $\mu$s, $7 \times 10^1$ $\mu$s, and $3 \times 10^{-2}$, respectively. The results shown in Fig. {\ref{fig3}}(c) were obtained from the quantum computer IBMQ Melbourne, where the average relaxation times $T_1$ and $T_2$ and readout error rate are about $6 \times 10^1$ $\mu$s, $6 \times 10^1$ $\mu$s, and $5 \times 10^{-2}$, respectively. The results shown in Fig. {\ref{fig4}} were obtained from the quantum computer IBMQ Sydney, where the average relaxation times $T_1$ and $T_2$ and readout error rate are about $1 \times 10^2$ $\mu$s, $8 \times 10^1$ $\mu$s, and $3 \times 10^{-2}$, respectively.
\section{Quantum phase estimation using the generalized $n$-qubit quantum circuits}
\begin{figure*}
\includegraphics{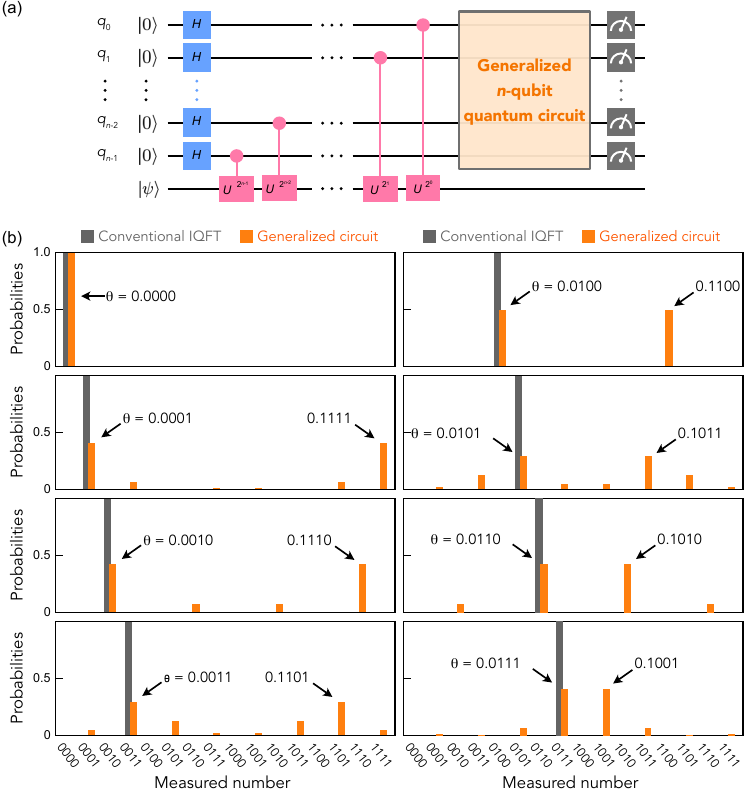}
\caption{\label{fig6}(a) A quantum circuit to perform the QPE using the generalized $n$-qubit quantum circuit. Compared to the conventional QPE algorithm, the IQFT part is replaced with the generalized $n$-qubit quantum circuit. The blue boxes and the red components represent the Hadamard gates and the controlled phase shift gates, respectively. (b) Probability distributions for the four-qubit QPE using the conventional IQFT (the gray bars) and the generalized quantum circuit (the orange bars) obtained from the exact calculations without errors. The generalized quantum circuit outputs binary representations of $2^n \theta$ and $2^n-2^n \theta$ with the maximum probability, where $\theta$ is the estimated phase for a unitary matrix $U$ and its eigen quantum state $\left|\phi\right>$ such that $U\left|\phi\right>=e^{2\pi i\theta} \left|\phi\right>$ (see Appendix C for detailed calculations of probability distributions).}
\end{figure*}

The generalized $n$-qubit quantum circuit shown in Fig. {\ref{fig3}}(b) was found to work well in the quantum phase estimation (QPE) problem{\cite{nielsen2010quantum,aspuru2005simulated}}. The QPE is a quantum algorithm to estimate the phase $\theta$ for a unitary matrix, $U$, and its eigen quantum state $\left|\phi\right>$ such that $U\left|\phi\right>=e^{2\pi i\theta} \left|\phi\right>$. The QPE is performed by the following three steps. Firstly, a product state $\left|0\right>\otimes \left|\phi\right>$ is prepared. Secondly, the Hadamard and controlled phase-shift gates are applied to the state as shown in Fig. {\ref{fig6}}(a), and the phase $\theta$ is introduced to the quantum state: $1/\sqrt{2^n} \sum_{k=0}^{2^n-1} e^{2\pi i\theta k} \left|k\right> \otimes \left|\phi\right>$, where $k$ denotes the integer representation of $n$-bit binary numbers. Thirdly, the IQFT is applied, and the state is converted into $\left|2^n \theta\right> \otimes \left|\phi\right>$, where we assume that $2^n \theta$ is an integer for simplicity. Then, we can obtain $\theta$ by measuring the quantum state. Here, we replace the IQFT with the generalized $n$-qubit quantum circuit shown in Fig. {\ref{fig3}}(b), and we obtain the final quantum state as $\sum_{j_1, \cdots, j_n \in \{0, 1\}}E_n^{0\oplus j_1}E_{n-1}^{j_1\oplus j_2}\cdots E_{1}^{j_{n-1}\oplus j_n}\left|j_n\cdots j_2 j_1\right>$, where $E_k^0=\left(1+e^{i\pi 2^{k}\theta}\right)/2$ and $E_k^1=\left(1-e^{i\pi 2^{k}\theta}\right)/2$, $\oplus$ symbol denotes the exclusive disjunction between two binary digits. The probability that the state $\left|j_n\cdots j_2 j_1\right>$ is observed is $\left|E_n^{0\oplus j_1}E_{n-1}^{j_1\oplus j_2}\cdots E_{1}^{j_{n-1}}\right|^2$. We analytically proved that the probability takes the maximum value when$j_n\cdots j_2 j_1$ is the binary representation of $2^n\theta$ or $2^n-2^n \theta$ [Fig. {\ref{fig6}}(b)]. The result means that the generalized $n$-qubit quantum circuit outputs the correct answer with the maximum probability at least in the QPE. Here, we note that the probability of obtaining the correct answer decreases exponentially with increasing the number of qubits. The generalized quantum circuits cannot solve the QPE in polynomial time with bounded error probability.
\begin{figure*}
\includegraphics{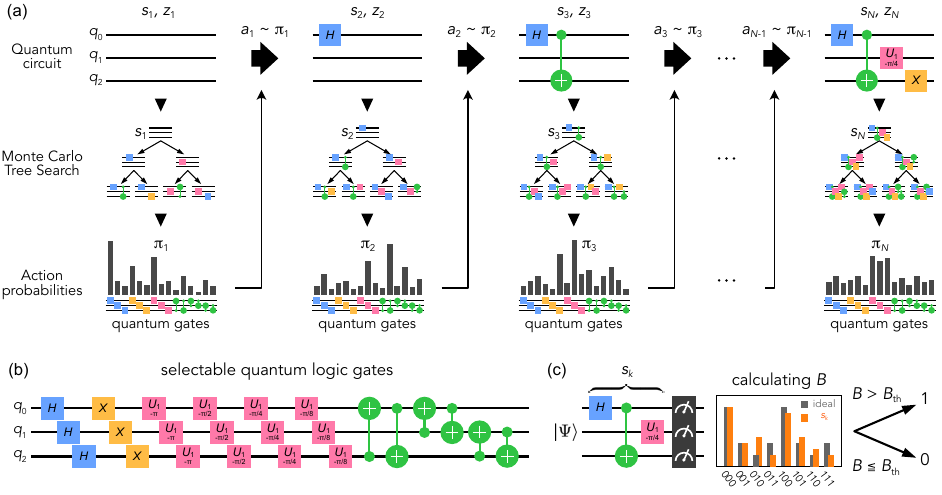}
\caption{\label{fig7}(a) The MCTS used as a policy of the reinforcement learning model. The MCTS searches quantum circuits by repeatedly appending a quantum gate starting from the present quantum circuit $s_k$. Reflecting state values of the searched quantum circuits, the MCTS generates a policy ${\boldsymbol{\pi}}_k$. Then, $s_k$ is updated to the next quantum circuit $s_{k+1}$ by selecting an action $a_k$ according to ${\boldsymbol{\pi}}_k$. By repeating to update the quantum circuit and conduct the MCTS, distilled quantum circuits are searched from a huge number of combinations of quantum gates. See Appendix D for more detailed procedure of the quantum circuit search. (b) An example of the set of quantum gates used in the reinforcement learning for 3-qubit quantum circuits. The yellow boxes and the green components represent the NOT gates and the controlled NOT gates, respectively. An action $a_k$ is defined as selecting a quantum gate from the gates in (b) and appending the gate to $s_k$. A policy ${\boldsymbol{\pi}}_k$ is defined as a probability distribution over the actions. (c) A reward calculation for a quantum circuit $s_k$. First, the Bhattacharyya coefficient $B$ between the ideal probabilities and output from $s_k$ is calculated for several initial quantum states $\left|\phi\right>$. Then, the reward $z_k$ is calculated as 1 or 0 depending on whether $B$ is larger than the threshold value $B_{\rm{th}}$ for all the initial quantum states or not.}
\end{figure*}
\begin{figure*}
\includegraphics{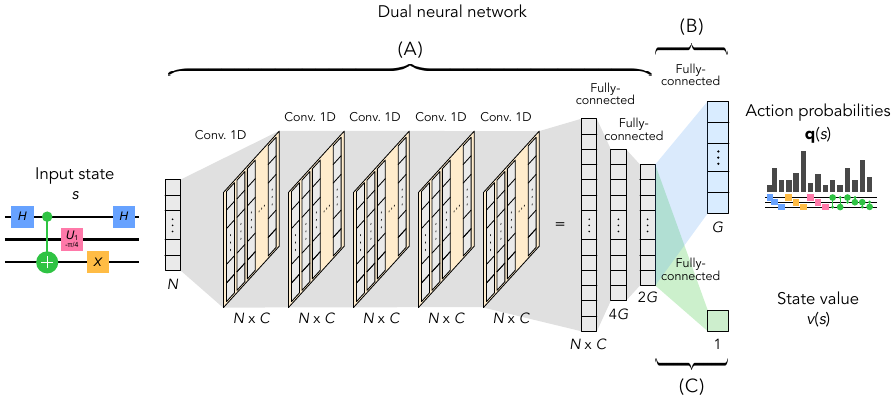}
\caption{\label{fig8}(a) The dual neural network consists of convolution (A), policy prediction (B), and state-value prediction (C) parts. The input to the network is an $N$-dimensional vector representing an input quantum circuit $s$. The outputs are a $G$-dimensional vector ${\textbf{q}}$ predicting a policy and a scalar $v$ predicting a state value of the input quantum circuit $s$, where $N$, $G$, and $C$ are the maximum length of input quantum circuits, the number of selectable quantum gates, and the number of channels, respectively. Details of the network structure and hyperparameters are summarized in Table {\ref{table1}}.}
\end{figure*}
\section{Reinforcement learning model}
We formulated the search problem as combination optimization of quantum logic gates. We have applied reinforcement learning {\cite{arulkumaran2017deep}} to the problem by treating quantum circuits and appending one quantum gate to the end of the quantum circuit as states and action, respectively [Fig. {\ref{fig7}}(a)]. A state $s_k$ consisting of $k$ quantum gates is updated to a state $s_{k+1}$ consisting of $k+1$ quantum gates by applying an action $a_k$. The quantum gates are selected from the Hadamard, NOT, Phase-shift, and CNOT gates, where magnitude of the phase shift is selected from $-2\pi /2^1$, $-2\pi /2^2$, $-2\pi /2^3$, or $-2\pi /2^4$ to construct a universal gate set. For example, we prepared 24 quantum gates for constructing 3-qubit quantum circuits as shown in Fig. {\ref{fig7}}(b). Then, various quantum circuits with different lengths and different combinations of quantum gates can be generated depending on a series of actions. $a_k$ is selected by using the action probability functions ${\boldsymbol{\pi}}(s_k)$ {\cite{kaelbling1996reinforcement}} obtained from the Monte-Carlo tree search (MCTS){\cite{silver2017mastering,silver2017mastering2}} (see Fig. {\ref{fig7}}). The component of the ${\boldsymbol{\pi}}(s_k)$ is defined as ${\boldsymbol{\pi}}(s_k, a) = {\rm{Pr}}(a|s_k)$, where $a$ is an action. The reward $z_k$ for the reinforcement learning is determined by calculating the Bhattacharyya coefficient $B_k$ between the output probabilities from $s_{k+1}$ and the correct answer [Fig. {\ref{fig7}}(c)]. When $B_k$ is above a threshold value, $B_{\rm{th}}$, for several input quantum states, we substitute 1 for $z_k$ and abort the search. When $B_k$ is less than $B_{\rm{th}}$, we assign 0 to $z_k$ and repeat searching for the next action, updating the state, and comparing $B_{k+1}$ and $B_{\rm{th}}$. If no states satisfy the condition until $k$ reaches the upper limit $N$, we assign -1 to $z_N$. $N$ determines the maximum length of the quantum circuits in the search, and we set $N$ to around $4n$ in this study. The total reward $z$ for the search is defined as the sum of the rewards $z=\sum_k z_k$ . The obtained set $(s_k, {\boldsymbol{\pi}}(s_k), z)$ is used as the training data to update the MCTS and dual neural network as explained below.\par
The reinforcement learning aims to learn a policy so as to maximize the total reward. As the policy model, we used the Alpha Zero algorithm based on the MCTS and a dual neural network{\cite{silver2017mastering,silver2017mastering2}}. The dual neural network predicts a policy, ${\bf{q}}$, and a state value, $v$, in response to the input state $s$ (see the next section, Fig. {\ref{fig8}}, and Table {\ref{table1}} for more details). The network parameters are trained by using the training data $(s, {\boldsymbol{\pi}}(s), z)$ by minimizing the loss function $L=(z-v)^2-{\boldsymbol{\pi}}^{\rm{T}} \log {\bf{q}}$. The MCTS efficiently searches quantum circuits reflecting the prediction of the dual neural network. The nodes and edges of the tree are defined as the states and the pairs of the state and action, respectively, where the root node is a present state, $s_k$ [see Fig. {\ref{fig7}}(a)]. Each edge $(s, a)$ has parameters: the visit count $N(s, a)$, the action value $Q(s, a)$, and the prior probability $P(s, a)$, where $N(s, a)$ and $Q(s, a)$ are initialized to zero. The tree search starts from the root node. Other nodes and edges are generated by selecting an action to maximize the upper confidence bound $Q(s,a)+U(s,a)$, where $U(s,a)=c_{\rm{puct}} P(s,a)\sqrt{\sum_b N(s,b)}/[1+N(s,a)]$, until it reaches one of the leaf nodes. $c_{\rm{puct}}$ is an exploration parameter. At the leaf node, the next node ($s'$) is appended and the prior probability $P(s', \cdot)$ and the state value $V(s')$ for $s'$ are predicted by using the dual neural network as $P(s',\cdot)={\bf{q}}(s')$ and $V(s')=v(s')$. Then the parameters of all visited edges are updated as follows: $N(s,a)\leftarrow N(s,a)+1$ and $Q(s,a)\leftarrow [N(s,a)Q(s,a)+V(s')]/[1+N(s,a)]$. By repeating the search and updating the parameters, we obtain the policy for the root state $s_k$: ${\boldsymbol{\pi}}(a|s_k)=N(s_k,a)/\sum_b N(s_k,b)$. Following the obtained policy ${\boldsymbol{\pi}}(s_k)$, the present state $s_k$ is updated to $s_{k+1}$ and we move on to the next search [see Fig. {\ref{fig7}}(a)].
\section{Dual neural network}
The dual neural network consists of convolution, policy prediction, and state-value prediction parts, respectively labeled as (A), (B), and (C) in Fig. {\ref{fig8}}. The convolution part (A) is composed of 5 convolution and 2 fully-connected layers with Leaky ReLU activation functions{\cite{xu2015empirical}}, the batch normalization technique{\cite{ioffe2015batch}}, and the dropout technique{\cite{srivastava2014dropout}}. The input to the convolution part is an $N$-dimensional vector representing a quantum circuit, and the output is $2G$-dimensional vectors. The policy prediction part (B) is a fully-connected network with a softmax activation function. The input to the policy prediction part is a $2G$-dimensional vector, and the output is a $G$-dimensional vector ${\bf{q}}$ predicting a policy to search quantum circuits, where $G$ is the number of selectable quantum gates. The state-value prediction part (C) is a fully-connected network with a hyperbolic tangent activation function. The input to the policy prediction part is a $2G$-dimensional vector, and the output is a scalar $v$ predicting a state value of the input quantum circuit. Please see Table {\ref{table1}} for more detailed parameters of the network. The network is trained by the adaptive moment estimation{\cite{kingma2014adam}} with the beta hyperparameters of $(0.9, 0.999)$ and the learning rate of 0.001.
%
%
\bibliography{aipsamp.bib}
\renewcommand{\thetable}{A·\Roman{table}}
\setcounter{table}{0}
\begin{table*}[hbtp]
\caption{Architecture of the dual neural network}
\label{table1}
\centering
\begin{tabular}{llll}
\hline
Part & Layer \# & Layer & Hyperparameters\\
\hline \hline
(A)&1&Convolution&output\_channel=256, kernel\_size=3, stride=1, padding=1\\
&2&Batch normalization& eps=1e-05, momentum=0.1, affine=True, track\_running\_stats=True\\
&3&Leaky ReLU&negative\_slope=0.01\\
&4&Convolution&output\_channel=256, kernel\_size=3, stride=1, padding=1\\
&5&Batch normalization& eps=1e-05, momentum=0.1, affine=True, track\_running\_stats=True\\
&6&Leaky ReLU&negative\_slope=0.01\\
&7&Convolution&output\_channel=256, kernel\_size=3, stride=1, padding=1\\
&8&Batch normalization& eps=1e-05, momentum=0.1, affine=True, track\_running\_stats=True\\
&9&Leaky ReLU&negative\_slope=0.01\\
&10&Convolution&output\_channel=256, kernel\_size=3, stride=1, padding=1\\
&11&Batch normalization& eps=1e-05, momentum=0.1, affine=True, track\_running\_stats=True\\
&12&Leaky ReLU&negative\_slope=0.01\\
&13&Convolution&output\_channel=256, kernel\_size=3, stride=1, padding=1\\
&14&Batch normalization& eps=1e-05, momentum=0.1, affine=True, track\_running\_stats=True\\
&15&Leaky ReLU&negative\_slope=0.01\\
&16&Fully-connected&in\_features=256×$G$ , out\_features=4×$G$, bias=True\\
&17&Batch normalization&eps=1e-05, momentum=0.1, affine=True, track\_running\_stats=True\\
&18&Leaky ReLU&negative\_slope=0.01\\
&19&dropout&p=0.3\\
&20&Fully-connected&in\_features=4×$G$ , out\_features=2×$G$, bias=True\\
&21&Batch normalization&eps=1e-05, momentum=0.1, affine=True, track\_running\_stats=True\\
&22&Leaky ReLU&negative\_slope=0.01\\
&23&dropout&p=0.3\\
\hline
(B)&24-1&Fully-connected&in\_features=2×$G$ , out\_features=$G$, bias=True\\
&25-1&Log softmax&\\
\hline
(C)&24-2&Fully-connected&in\_features=2×$G$ , out\_features=1, bias=True\\
&25-2&tanh&\\
\hline
\end{tabular}
\end{table*}
\end{document}